\documentclass[twocolumn,aps,floats,superscriptaddress,showpacs]{revtex4}
\usepackage{amsmath}
\usepackage{graphicx}

\newcommand{\mr}[1]{{{\mathrm{#1}}}}
\newcommand{\mcal}[1]{{\mathcal{#1}}}
\newcommand{\hs}{\hspace{-0.10cm}}
\newcommand{\pdagger}{{\phantom{\dagger}}}
\newcommand{\dt}{\partial_\tau}
\newcommand{\inte}{\int_0^\beta \!\!\!\! \mr{d}\tau}
\newcommand{\w}{\omega}
\newcommand{\fsd}{f^{\dagger}_{\sigma}}
\newcommand{\fs}{f_{\sigma}}
\newcommand{\s}{\sigma}
\newcommand{\qbar}{\overline{Q}}

\begin{document}

\title{Climbing the Entropy Barrier: Driving the Single-
towards the Multichannel Kondo Effect by a Weak Coulomb
Blockade of the Leads}


\author{S. Florens}
\affiliation{\mbox{Institut f\"ur Theorie der Kondensierten Materie, Universit\"at
Karlsruhe, 76128 Karlsruhe, Germany}}
\author{A. Rosch}
\affiliation{\mbox{Institut f\"ur Theorie der Kondensierten Materie, Universit\"at
Karlsruhe, 76128 Karlsruhe, Germany}}
\affiliation{Institut f\"ur Theoretische Physik, Universit\"at
zu K\"oln, 50937 K\"oln, Germany}

\begin{abstract}
  We study a model proposed recently
  in which a small quantum dot is coupled symmetrically to several
  large quantum dots characterized by a charging energy $E_c$. Even if
  $E_c$ is much smaller than the Kondo temperature $ T_K$, the
  long-ranged interactions destabilize the single-channel Kondo effect
  and induce a flow towards a multi-channel Kondo fixed point
  associated with a {\em rise} of the impurity entropy with decreasing
  temperature. Such an ``uphill flow'' implies a {\em negative} impurity
  specific heat, in contrast to all systems with local interactions.
  An exact solution found for a large number of channels allows us to
  capture this physics and to predict transport properties.
\end{abstract}


\maketitle

Simple models of non-Fermi liquids, such as the multi-channel Kondo
model, have attracted much theoretical interest \cite{cox}, especially due to
their striking properties like zero-bias anomalies or a finite entropy
at zero temperature, $S(T=0)=\ln(g)$, with a non-integer $g$
\cite{natan}.  So far, however, their experimental realization has been
  a challenging task.  The advent of nanostructures that can be
  designed and tuned more easily than solid state systems is a major step towards observing these interesting
  strong-correlation effects in the laboratory.
  
  The multi-channel Kondo model describes a single spin coupled
  symmetrically to several {\em independently} conserved conduction electron
  ``channels''.  Recently Oreg and Goldhaber-Gordon \cite{oreg}
  suggested that these channels can be realized by attaching several
  large quantum dots serving as ``leads'' to a single small dot.  They
  pointed out that a sufficiently large charging energy, $E_c \gtrsim
  T_K$, in those leads can supress all low-energy cotunneling
  processes between them which would otherwise mix the channels
  and destroy the multi-channel physics. The experimental realization
  of such a system is a demanding task as the size of the large
  dots has to be chosen such that the level-splitting $\Delta_L$ is
  sufficiently small and, simultaneously, the charging energy is
  large enough, $\Delta_L \ll T_K^{\text{multi}} \lesssim T_K
  \lesssim E_c$, where $T_K^{\text{multi}}$ is the Kondo temperature
  of the multi-channel Kondo model and $T_K$ refers to the
  single-channel Kondo temperature for $E_c=0$. Furthermore,
  considerable fine-tuning, using gate voltages, is required
  to guarantee that all leads couple equally  to the spin
  residing on the small dot. Pustilnik {\it et al.} \cite{pustilnik}
  have recently discussed in detail how to achieve such a fine-tuning 
  by calculating the conductance for small variations in the
  coupling to the various channels. The idea that interactions in the
  leads can lead to multi-channel physics was also put forward earlier
  by Coleman and Tsvelik \cite{coleman}, 
  and received recent additional support \cite{anders}.

  What happens if the charging energy $E_c$ of the leads is much
  smaller than the Kondo temperature, $E_c \ll T_K$? In this case a
  single-channel Kondo resonance will develop upon lowering $T$.
  According to conventional wisdom, the Fermi-liquid fixed point of
  the single-channel Kondo effect can never be destroyed by small
  perturbations (for example a weak magnetic field, $B\ll T_K$, does
  not prevent the formation of the Kondo resonance).  Indeed, the
  powerful ``$g$-theorem'' of Affleck and
  Ludwig~\cite{affleck,gtheorem} proves that no small local
  perturbation can destabilize such a zero-entropy fixed point: The
  impurity entropy $\ln(g)$ of boundary conformal field theories (to
  which Kondo models belong) always {\em decreases} under
  renormalization group flow.  According to this theorem, a flow from
  the multi-channel to the single channel fixed point is possible, but
  not vice versa.  However, the $g$-theorem does not cover the
  situation under discussion, where long-range interactions induce a
  Coulomb blockade in the leads.

The purpose of this Letter is to show that in such a situation tiny
charging energies $E_c\ll T_K$ can destroy the single-channel Kondo
effect and stabilize a multi-channel fixed point. As the multi-channel
system is characterized by a {\em finite} residual entropy
[$S(T=0)=\ln \sqrt{2}$ for two channels] this requires a negative
impurity specific heat in some $T$ range. A simple argument in
favor of such a scenario is that the existence of the single-channel
Kondo effect implies resonant tunneling between the leads. The Coulomb
blockade in the leads, however, prohibits such a  resonance. The
Kondo effect cannot overcome the Coulomb blockade, as resonant
tunneling in an energy window of width $T_K$ costs a charging energy
of order $E_c \langle (\Delta N)^2 \rangle \sim E_c
\frac{T_K}{\Delta_L}$, where $\Delta N$ are charge fluctuations
induced by the resonant tunneling and $\Delta_L$ is the level spacing
in the leads.  These energy costs are larger than $T_K$ whenever $E_c
> \Delta_L$, suggesting that for $E_c \gg \Delta_L$ a
multi-channel Kondo effect will form.

We consider a model where the spin on the ``small'' quantum dot
(represented by Abrikosov fermions $\fsd$) couples to several larger
dots displaying charging effects:
\begin{eqnarray}
\nonumber
H &=& \sum_{k\s\alpha} \epsilon_k  c^\dagger_{k\s\alpha} c_{k\s\alpha} + 
\sum_{\alpha} E_c \Big[ \sum_{k\s} : c^\dagger_{k\s\alpha} c_{k\s\alpha}\!:\Big]^2 \\
\label{eq:hamiltonian}
&  &+ \frac{J}{N K} \sum_{k,k'}\, \, \sum_{\s,\s'=1}^N \, \, \sum_{\alpha,\alpha'=1}^K \!\! f^\dagger_{\s'} \fs c^\dagger_{k\s\alpha}
c_{k'\s'\alpha'}.
\end{eqnarray}
where $:\ldots:$ denotes normal ordering and we assume non-degenerate
charge states in the leads.  In view of greater generality that will
be useful in the following, we suppose an arbitrary number $K$ of
interacting leads, and we consider an $SU(N)$ spin in a representation
which fulfills the constraint $\sum_{\s=1}^N \fsd \fs = N/2$. For
$N=2$, the last term in Eq.~(\ref{eq:hamiltonian}) describes (up to a
potential scattering term) the usual exchange coupling of a spin $1/2$
to a symmetric combination of electrons from all leads,
$c_{k\sigma,s}^\dagger=\frac{1}{\sqrt{K}}\sum_{\alpha=1}^K
c^\dagger_{k\sigma \alpha}$. Therefore the conventional one-channel
Kondo effect develops for $E_c=0$.  Note the
symmetric coupling of the spin to all leads; as mentioned above, this
requires fine-tuning in experiments.

\begin{figure}
\begin{center}
\includegraphics[width=0.84 \linewidth]{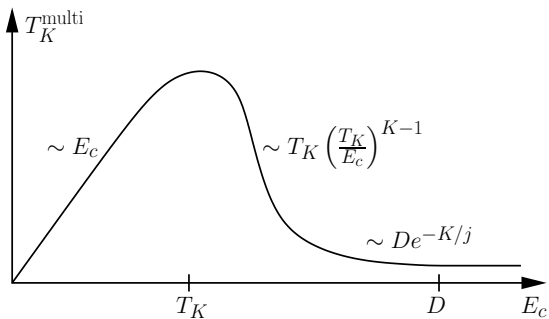}
\end{center}
\caption{Schematic plot of the multi-channel Kondo temperature $T_K^\mr{multi}$ 
versus $E_c$. Note the maximum for $E_c \sim T_K$.}
\label{fig:TKm1}
\end{figure}

In order to make contact with previous results \cite{oreg,pustilnik},
we first use the perturbative renormalization group (RG) to discuss the
case of large charging energy $T_K \ll E_c \ll D$, where $D$ is an
ultra-violet cutoff set by the bandwidth or the charging energies in
the small dot. For cutoffs large compared to $E_c$, the Coulomb
blockade does not modify the RG flow for the dimensionless coupling
$j=J N_f$ ($N_f$ is the density of states).  For a running cutoff
$\Lambda \gg E_c$, one therefore finds within one-loop RG $\partial
j/\partial \ln \Lambda = - j^2$ and at the scale $\Lambda=E_c$ the
running $j(\Lambda)$ takes the value $j(E_c)=1/\ln[E_c/T_K]$ with
$T_K=D e^{-1/j}$ \cite{twoloop} and $j(E_c)\ll 1$ as $E_c \gg T_K$.
For $\Lambda \ll E_c$ the charge on each lead is conserved separately
and all processes which mix channels are frozen out. Therefore one
expects a flow \cite{cox} to the $K$-channel Kondo fixed point with
$\partial j/\partial \ln \Lambda = - j^2/K$ with the flow starting at
$j(E_c)$. From this we can read off the multi-channel Kondo
temperature \cite{twoloop} for $T_K \ll E_c \ll D$
\begin{eqnarray}\label{tkMulti}
T_K^{\text{multi}}\approx E_c e^{-K/j(E_c)} 
\approx T_K \, \left(T_K/E_c\right)^{K-1}.
\end{eqnarray}
For $E_c>D$, the multichannel equation determines the  RG flow alone as
channel mixing is suppressed, which leads to the usual result
$T_{K}^\mr{multi} = D e^{-K/j}$. Fig.~\ref{fig:TKm1} shows the enhancement
of  $T_K^\mr{multi}$  with decreasing $E_c$.

To analyze the physics at small $E_c$, we introduce a phase 
(or slave rotor) representation \cite{phase,serge} of the charging energy:
\begin{eqnarray}
\nonumber
H &=& \sum_{k\s\alpha} \epsilon_k a^\dagger_{k\s\alpha} a^\pdagger_{k\s\alpha} 
+ \sum_{\alpha} E_c \widehat{L}_\alpha^2 \\
\label{eq:rotor}
&+& \frac{J}{N K}  \sum_{kk'\s\s'\alpha\alpha'} \!\!\!\!
f^\dagger_{\s'} \fs a^\dagger_{k\s\alpha}
a^\pdagger_{k'\s'\alpha'} e^{i\theta_\alpha - i\theta_{\alpha'}},
\end{eqnarray}
where we have set $c^\dagger_{k\s\alpha} \hs=\hs a^\dagger_{k\s\alpha}
\exp(i\theta_\alpha)$ and $L_\alpha \hs=\hs -i
\partial/\partial\theta_\alpha$.  
This effective {\em matrix} of Kondo couplings $J_{\alpha\alpha'}(\theta) = 
J \exp(i\theta_\alpha - i\theta_{\alpha'})$ allows to reinterpret the original
idea of Oreg and Goldhaber-Gordon. At large $E_c$ the phases $\theta_\alpha$
fluctuate wildly in $J_{\alpha\alpha'}(\theta)$ for
$\alpha\neq\alpha'$, which leads to a flow towards a diagonal coupling
at low energy. In this case, channel number is conserved and a
multi-channel Kondo effect will develop. However, for $E_c$ equal to
zero strictly, the phases are locked and drop from the Kondo coupling
in Eq.~(\ref{eq:rotor}), leaving a model displaying Fermi liquid
properties and complete screening of the spin.

To study the effects of a small $E_c\ll T_K$ within perturbation theory, we
expand Eq.~(\ref{eq:rotor}) around constant $\theta_\alpha$. This
expansion is valid as long as $\langle
(\theta_\alpha(\tau)-\theta_\alpha(\tau'))^2 \rangle \ll 1$, which
allows us to reliably calculate the correction $\Delta F$ to the free energy for
$T\gg E_c$. After some straightforward but tedious manipulations we
obtained 
\begin{eqnarray}
 \label{Fpert}
\frac{\Delta F}{N} &\approx & -\frac{K-1}{2 K} \inte \left\langle
[\theta(\tau) - \theta(0)]^2\right\rangle T(\tau) G_0(\tau)\\
&\approx &  \frac{K-1}{\pi^2K} \int_T^{T_K} \frac{2 E_c}{\w^2} \w \approx  
2 \frac{K-1}{\pi^2K} E_c \log\!\left[\frac{T_K}{T}\right],\nonumber
\end{eqnarray}
where $T(\tau)$ is the $T$-matrix of the symmetric channel and 
$G_0(i\w_n) = \sum_k 1/(i\w_n-\epsilon_k)$ the bare local Green's function of 
the conduction electrons. For the last equality in
(\ref{Fpert}) we used that $\pi N_f \text{Im} T(\omega)=1$ for $\w,T \ll
T_K$ which follows from  Friedel's sum rule. It is therefore valid for $E_c \ll T
\ll T_K$.  Up to (known) prefactors of order $1$, the impurity entropy
in this regime is thus given by
\begin{eqnarray} \label{sImpAna}
S(T)/N \approx T/T_K + E_c/T,
\end{eqnarray}
where the first term arises from the usual Kondo effect and the second
from the charge fluctuations described by Eq.~(\ref{Fpert}). This
proves that the impurity entropy will show a minimum at the scale $T_S
= \sqrt{E_c T_K}$ and implies a {\em negative} impurity specific heat
for $E_c \ll T \ll T_S$ (the specific heat of the total system $\sim
T/\Delta_L$ remains positive as we assumed a neglegible level spacing
$\Delta_L \ll E_c$).  Moreover, the entropy per spin reaches values of
order $1$ for $T \sim E_c$ where the expansion breaks down, opening
the possibility of a flow towards a multi-channel fixed point. When we
take into account that the crossover at $E_c$ takes place deep in the
strong-coupling regime, this suggests that the corresponding
multi-channel Kondo temperature is directly given by $E_c$, as shown
schematically in Fig.~\ref{fig:TKm1}.

To be able to describe the crossover at $E_c$ and the physics for $T
\lesssim E_c \ll T_K$ we need a non-perturbative method. We have found
an exactly solvable limit of the Hamiltonian~(\ref{eq:hamiltonian})
that confirms the previous calculations, and 
also offers a direct computational tool to describe the crossover from single-channel 
to multi-channel physics as a function of $T$ and charging energy. The idea is to 
solve the problem by taking both $N$ and $K$ to be large (we recall that we 
have considered a generalized model with an $SU(N)$ quantum spin in the 
dot and $K \equiv \gamma N$ interacting leads coupled to it). The technical step 
is to notice that the Kondo interaction in~(\ref{eq:rotor}) can be decoupled using 
a {\it single} bosonic field $B(\tau)$ conjugate to $\sum_{k\s\alpha} \fsd a_{k\s\alpha} 
\exp(-i\theta_\alpha)$. Integrating out the leads, we obtain the following action 
in imaginary time:
\begin{eqnarray}
\nonumber
\mcal{S} \!\!&=& \!\! \inte \frac{K B^\dagger B}{J} + 
\sum_\alpha \frac{(\dt \theta_\alpha)^2}{4 E_c} +
\sum_\s \fsd (\dt - \mu) \fs + \mu \frac{N}{2}\\
\label{eq:action}
&\!\!\!\!\!\!\!\! + & \!\!\!\!\!  \inte \!\! \inte' \; \frac{G_0(\tau-\tau')}{N} \sum_{\s\alpha}
[\fsd B e^{-i\theta_\alpha}]_\tau [\fs B^\dagger e^{i\theta_\alpha}]_{\tau'},
\end{eqnarray}
where $\mu$ is a complex Lagrange multiplier used to enforce the
constraint on the spin size. We can introduce
\cite{olivier} two fields $Q(\tau,\tau')$ and
$\qbar(\tau,\tau')$ to decouple fermions from bosons in the last term
of~(\ref{eq:action}). Then the $\fsd$
and $\theta_\alpha$ variables are integrated out to obtain an effective action
$\mcal{S}[B,Q,\qbar,\mu]$ which is proportional to $N$ and therefore
solved by a saddle-point when $N\rightarrow \infty$. Using
time-translational invariance and particle-hole symmetry (so that
$\mu=0$), we obtain the self-consistent equations 
\begin{eqnarray}
\label{eq:Gf}
G_f(i\w_n) &\equiv& \big<f^\dagger(i\w_n) f(i\w_n)\big> 
= \frac{1}{i\w_n - B^2 Q(i\w_n)} \\
G_X(\tau) &\equiv& \big<e^{i\theta(\tau)-i\theta(0)}\big>\\
Q(\tau) &=& \gamma G_0(\tau) G_X(\tau) \\
\qbar(\tau) &=& - B^2 G_0(\tau) G_f(\tau) \\
\label{eq:J}
\frac{1}{J} &=&  \inte \; G_0(\tau) G_X(\tau) G_f(\tau) 
\end{eqnarray}
where the condensate $B$ is determined from Eq.~(\ref{eq:J}) and
the correlator $G_X(\tau)$ is computed from the action
\begin{equation}
\label{eq:phase}
\mcal{S} = \inte \frac{(\dt \theta)^2}{4 E_c} + \inte \! \inte' \; 
\qbar(\tau-\tau')
\; e^{i\theta(\tau)-i\theta(\tau')}.
\end{equation}
In principle, the model~(\ref{eq:phase}) can be solved very
efficiently by Monte Carlo \cite{herrero}, but we have chosen 
to simplify the numerics by using the spherical limit as an 
approximation to this rotor model.
This is done by introducing a further large $M$ expansion, where we
generalize $e^{i\theta}\equiv X$ with the constraint $|X|^2=1$ to $M$
fields $X_{i}$ with $\sum_{i=1}^M |X_{i}|^2=M$ (see Ref.~\cite{serge}
for details). We then obtain 
\begin{eqnarray}
\label{eq:GX}
G_X^{-1}(i\nu_n) &=& \nu_n^2/(4E_c) + \lambda + \qbar(i\nu_n)\\
\label{eq:lambda}
G_X(\tau=0) &=& 1.
\end{eqnarray}
The parameter $\lambda$ is determined from Eq.~(\ref{eq:lambda}), reflecting the 
constraint $\big|e^{i\theta}\big|^2=1 $ in average.

\begin{figure}
\begin{center}
\includegraphics[width=0.90 \linewidth]{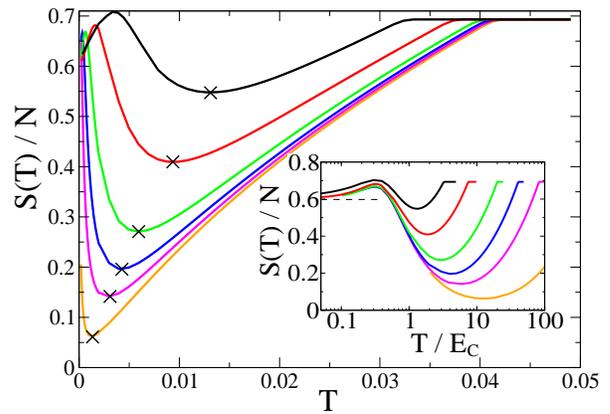}
\end{center}
\caption{Entropy $S(T)$ for $T_K/E_c =400$, $80$, $40$, $20$, $8$, $4$
  (bottom to top) with $J N_f=1/4$, $\gamma=1$ and a flat density of
  states of half-width $D=2$ (giving $T_K\approx 0.04$). The minimum
  of $S$ is located at $T_S \approx 0.7 \sqrt{E_c T_K}$ (crosses). For
  $T\ll T_S$ and $E_c \ll T_K$, the entropy is a function of $T/E_c$
  only (see inset), which shows that $E_c$ can be identified with the
  multi-channel Kondo temperature $T_K^{\text{multi}}$ in this limit.
  The dotted line in the inset denotes the exact value for $S(T=0)$
  taken from Ref.~\cite{olivier}. Note that in the regime $0.4 E_c <T <T_S$,
  the impurity specific heat $C=T dS/dT$ is {\em negative}.}
\label{fig:entropy2}
\end{figure}

We first investigate the general properties of the system of
equations~(\ref{eq:Gf}-\ref{eq:J}) and (\ref{eq:GX}-\ref{eq:lambda}).
For $E_c=0$ we recover nicely the usual large-$N$ limit of the
single-channel Kondo model \cite{coleman2}, since $G_X(\tau) = 1$
follows from Eqs.~(\ref{eq:GX}-\ref{eq:lambda}) in this case. For $E_c
\neq 0$, this behavior  also holds at $E_c \ll T$, reflecting the fact
that the single-channel Kondo fixed point controls the regime $E_c \ll
T \ll T_K$. However, at $T=0$ a low-frequency
analysis of our integral equations \cite{olivier,serge} shows the
appearance of universal power laws characteristic of the
multi-channel fixed point \cite{olivier}, $G_f(i\w) \sim \frac{1}{i
  \w} |\w|^{1/(1+\gamma)}$ and $G_X\sim 1/ |\w|^{1/(1+\gamma)}$.

Our equations allow also to calculate physical quantities 
at intermediate coupling.
Fig.~\ref{fig:entropy2} shows the impurity entropy, i.e. the
difference of the total entropy and the entropy in the absence of the
spin. As predicted by Eq.~(\ref{sImpAna}), there is a minimum in
$S(T)$ at the scale $T_S\simeq\sqrt{E_c T_K}$ and the specific heat is
negative for $E_c \lesssim T <T_S$.  Close to the low-$T$ fixed point
the specific heat is positive. Accordingly, the entropy drops below
$E_c$ and for $T\to 0$  reaches the finite value characteristic for the
$K$-channel fixed point \cite{olivier}.  The scaling plot in the inset
of Fig.~\ref{fig:entropy2} shows that $E_c$ can be identified with the
multi-channel Kondo temperature $T_K^{\text{multi}}$ if $E_c \ll T_K$,
as discussed above.
For $E_c > T_K$, we find only the free-spin solution with $B=0$ and
no multi-channel physics. This is consistent with Eq.~(\ref{tkMulti}),
which shows that $T_K^{\text{multi}}$ drops rapidly to $0$ for $K\to
\infty$ if $E_c>T_K$. For finite $K$ and low $T$, multi-channel behavior is of course 
maintained for arbitrarily large $E_c$.

\begin{figure}
\begin{center}
\includegraphics[height=5.2cm]{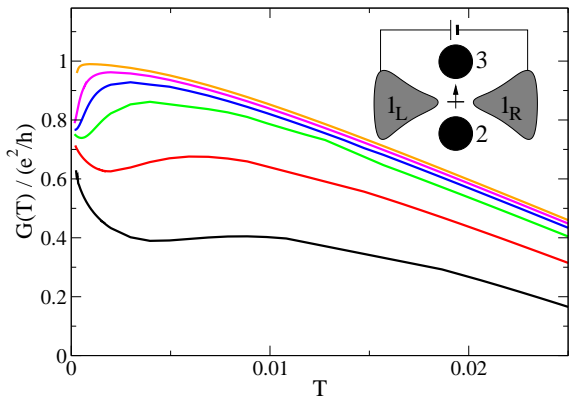}
\end{center}
\caption{
Conductance $G(T)$ for $E_c$ = 0.0001, 0.0005, 0.001, 0.002, 0.005, 
  0.01 (top to bottom) as in Fig.~\ref{fig:entropy2}.  For $T\to 0$
  and $\gamma=1$, $G$ approaches the universal value $G=\pi G_0/4$. In
  contrast to the entropy (inset of Fig.~\ref{fig:entropy2}), large
  non-universal corrections spoil the one-parameter scaling with
  $T/E_c$. 
  Inset: Experimental setup suggested in Ref.~\cite{oreg}.  Two leads
  ($1_L$ and $1_R$) are connected to a voltage source, the other
  ``leads'' $2,3$ are large quantum dots with a charging energy $E_c$
  and small level spacing $\Delta_L\ll E_c$, prohibiting charge 
  transport between $1$ and $2,3$.  Therefore, a mapping onto our
  Hamiltonian~(\ref{eq:hamiltonian}) with $K=3$ is possible, where
  channel $1$ arises from the even combination of electrons
  \cite{oreg,pustilnik} in $1_L$ and $1_R$. However, fine-tuning is
  required~\cite{pustilnik} to obtain symmetric coupling to all leads
  at lowest energies.}
\label{fig:conductance3}
\end{figure}
Although this thermodynamic analysis provides interesting insights
into the model considered here and the general question of how
long-range interactions can destabilize zero-entropy fixed points, it is
almost impossible to measure the entropy of quantum dots. Therefore
we have also calculated the conductance through the dot using a setup
suggested by Oreg and Goldhaber-Gordon \cite{oreg}, which is sketched
in Fig.~\ref{fig:conductance3}. In a situation where the odd combination 
of electrons in leads $1_L$ and $1_R$ decouples from the dot, the linear
conductance can be calculated from the imaginary part of the $T$-matrix 
\cite{meir,pustilnik}:
\begin{eqnarray}
\label{eq:landauer}
G &=& \frac{N e^2}{2 K \pi \hbar} \int \mr{d}\w
\frac{\partial n_F(\w)}{\partial \w}
 \mr{Im} G_0(\w) \mr{Im} T(\w)\\
T(\tau) &=&  B^2 G_X(\tau) G_f(\tau).
\end{eqnarray}
For $E_c \ll T \ll T_K$, the single-channel Kondo effect results in
resonant scattering among the $K$ equivalent leads and therefore in a
conductance $G \approx (N/K) G_0=(1/\gamma) G_0$ with $G_0=e^2/(2 \pi
\hbar)$. For $T\ll E_c\approx T_K^{\text{multi}}$, the conductance is
governed by the multi-channel $T$-matrix which has been calculated
from conformal field theory by Parcollet {\it et al.} \cite{olivier}
for arbitrary $K$ and $N$. For $K=N=2$ one obtains $G=G_0$, while
 in the large $N$ limit one gets $G=G_0 \; \pi/(2+2\gamma) 
\tan\!\left[\pi/(2+2\gamma)\right]$.

We believe that all the results obtained in the previous large $N$ and
$K$ limit are qualitatively valid for the experimentally relevant case
$N=2$ and $K=2$ or $3$. First, this is indicated by our perturbative
expansion (\ref{Fpert}) for $T\gg E_c$ and $E_c \ll T_K$ which proves
the existence of the minimum in $S(T)$ for {\it arbitrary} values of
$N$ and $K$.  Second, the Coulomb blockade of the leads makes the
multi-channel fixed point obviously stable against inter-lead
tunneling. Finally, we have verified our scenario for $K=2$ in  a
strong-coupling expansion of the Hamiltonian (\ref{eq:hamiltonian}),
taking first the limit $J \to \infty$ (and therefore $T_K \to \infty$)
and analyzing the resulting model for large $E_c \ll J$. In this limit
we recover a 2-channel Anderson model which can be mapped via a
Schrieffer-Wolff transformation to the two-channel Kondo model.

In conclusion, we have shown that tiny charging energies in the leads,
$E_c \ll T_K$, can destabilize the single-channel Kondo effect and
induce a flow towards the multi-channel Kondo fixed point. While in
systems with local interactions the $g$-theorem \cite{affleck,gtheorem}
guarantees that the impurity entropy always decreases with decreasing
$T$,  in our case it will rise for $E_c \lesssim T \lesssim
\sqrt{E_c T_K}$. Our observation that the multi-channel Kondo fixed
point can be stabilized even by small charging energies should help to
realize the experimental setup proposed by Oreg and Goldhaber-Gordon
\cite{oreg}. To obtain a high multi-channel Kondo temperature
$T_K^{\text{multi}}$, parameters with $E_c\approx T_K$ seem to be the
most promising candidate (see Fig.~\ref{fig:TKm1}).

We acknowledge useful discussions with N.~Andrei, L.~Borda, J.~von
Delft, Y.~Oreg, M.~Vojta and P.~W\"olfle and financial support from
the DFG by the Emmy-Noether program (A.R.) and the Center for
Functional Nanostructures (S.F.). Part of this work was performed
at the LMU M\"unchen (A.R.).


\begin{thebibliography}{5}
\bibitem{twoloop} To obtain the correct pre-exponential factors, a
  two-loop analysis is required, see e.g. Ref.~\cite{cox}.
\end{thebibliography}
\end{document}